# Scattering and Pairing in Cuprate Superconductors

**Louis Taillefer**

Canadian Institute for Advanced Research
Regroupement Québécois sur les Matériaux de Pointe
Département de physique, Université de Sherbrooke, Sherbrooke, Canada

E-mail: Louis.Taillefer@USherbrooke.ca

**Abstract**

The origin of the exceptionally strong superconductivity of cuprates remains a subject of debate after more than two decades of investigation. Here we follow a new lead: The onset temperature for superconductivity scales with the strength of the anomalous normal-state scattering that makes the resistivity linear in temperature. The same correlation between linear resistivity and $T_c$ is found in organic superconductors, for which pairing is known to come from fluctuations of a nearby antiferromagnetic phase, and in pnictide superconductors, for which an antiferromagnetic scenario is also likely. In the cuprates, the question is whether the pseudogap phase plays the corresponding role, with its fluctuations responsible for pairing and scattering. We review recent studies that shed light on this phase – its boundary, its quantum critical point, and its broken symmetries. The emerging picture is that of a phase with spin-density-wave order and fluctuations, in broad analogy with organic, pnictide, and heavy-fermion superconductors.

## 1. INTRODUCTION

Superconductivity is a fascinating, almost magical, property of matter. The ability of a metal to undergo a phase transition and enter a new state of matter in which electrons carry electricity perfectly, with infinite conductivity, sounds like utopia or mathematical fancy. Yet many real materials (such as aluminum, lead, and tin) do have this property of superconductivity. Unfortunately, they do so only at extremely low temperature, near absolute zero. Many modern-day alchemists have dreamt of finding a material with a superconductivity that could survive up to room temperature, at which the wonders of this unique quantum-mechanical state could be exploited more easily. This dream was fueled by the discovery of cuprates in 1986 [1], a family of copper oxide materials in which superconductivity has been found to persist as high as 164 K – halfway to room temperature.

**Cuprate:** copper oxide material made of layers of $CuO_2$ in which superconductivity occurs upon either hole or electron doping, with $T_c$ values as high as 164 K or 25 K, respectively

Having stimulated over 100,000 publications, the question of what causes superconductivity in the cuprates is widely considered to be one of the great challenges of condensed-matter physics [2]. On the twentieth anniversary of its discovery, it was deemed "a mystery that defies solution" [3]. Since then, however, exciting developments have given researchers hope that a solution may in fact be within reach [4]. In particular, a new family of superconductors – the pnictides – was discovered [5], with critical temperatures as high as 57 K [6, 7]. In this review, I discuss some of the recent developments that shed light on the two major questions of cuprate superconductivity: What causes electron pairing? What is the nature of the pseudogap phase, the enigmatic region of the phase diagram that overlaps with much of the superconducting phase? I do not attempt to review the field but rather explore a particular perspective, based on the conjecture that the pseudogap phase is fundamentally a phase with spin-density-wave (SDW) order, ending at a quantum critical point (QCP), with its fluctuations dominating the scattering of electrons and their pairing. This would make cuprates similar to heavy-fermion, organic, and pnictide superconductors, for which superconductivity is typically found in close proximity to SDW order. While many researchers believe that distinct theories are required for the different families of superconductors, I shall argue here that a fundamentally similar pairing mechanism operates, and this common perspective suggests that a solution may be within reach.

**Pnictide:** iron-based material made of layers of $Fe_2As_2$ in which superconductivity can occur, with $T_c$ values as high as 57 K

**Pseudogap phase:** enigmatic region of the cuprate phase diagram delineated by a crossover temperature $T^*$ below which the electronic density of states is partially gapped

**Spin density wave (SDW):** antiferromagnetic modulation of the spin density in a metal; the new periodicity causes a reconstruction of the Fermi surface such that a large hole surface is typically transformed into small electron and hole pockets

**Quantum critical point (QCP):** point at $T = 0$ in the phase diagram of a material where an ordered phase ends, as a function of pressure, doping or magnetic field; antiferromagnetic QCPs are found in heavy-fermion, organic, cuprate and pnictide superconductors

## 2. PHASE DIAGRAM

The doping phase diagram of hole-doped cuprates is sketched in **Figure 1**. With increased hole concentration (doping) $p$, the materials go from being antiferromagnetic insulators at zero doping to metals at high doping. Given their density of one electron per Cu in the undoped state, they should be metals even at $p = 0$, with a Fermi surface volume containing $1 + p$ holes, but strong on-site repulsion prevents electron motion and turns the material into a Mott insulator at low doping. At intermediate doping, between the insulator and the metal, there is a central region of superconductivity, delineated by a critical temperature $T_c$, which can rise to values of order 100 K – higher than in any other family of materials. Near optimal doping, the normal state above $T_c$ is referred to as a "strange metal", characterized by a resistivity that is linear in temperature. In the midst of this strange-metal region, the pseudogap phase sets in, below a crossover temperature $T^*$ at which most physical properties undergo a significant change [8]. The question is whether the pseudogap phase is a precursor to some "hidden" ordered state with broken symmetry or a precursor to the Mott insulator, with no broken symmetry.

To explore this landscape, we shall start from the far right side of **Figure 1**, in the overdoped metallic state. This state is characterized by a large Fermi surface whose volume contains $1 + p$ holes per Cu atom, as determined by angle-dependent magneto-resistance (ADMR) [9], angle-resolved photoemission spectroscopy (ARPES) [10] and quantum oscillations [11], all performed on the single-layer cuprate $Tl_2Ba_2CuO_{6+\delta}$ (Tl-2201). The low-temperature Hall coefficient $R_H$ of overdoped Tl-2201 is positive and equal to $1 / e (1 + p)$ [12], as expected for a single-band metal with a hole density $n = 1 + p$. Conduction in the normal state obeys the Wiedemann-Franz law [13], a hallmark of Fermi-liquid theory. At the highest doping, beyond the superconducting phase (**Figure 1**), the electrical resistivity $\rho(T)$ of Tl-2201 exhibits the standard $T^2$ temperature dependence of a Fermi liquid [14], also observed in $La_{2-x}Sr_xCuO_4$ (LSCO) [15].

---

**Fermi surface:** boundary in $k$-space that separates occupied electron states from unoccupied states; its volume is directly proportional to the carrier density; when closed, it can be electron-like (enclosing occupied states) or hole-like (enclosing unoccupied states)

**ADMR:** angle-dependent magneto-resistance

**ARPES:** angle-resolved photo-emission spectroscopy

**Tl-2201:** $Tl_2Ba_2CuO_{6+\delta}$

**Quantum oscillations:** oscillations in the resistance or magnetization of a metal as a function of magnetic field $B$ that result from cyclotron motion and Landau quantization of energy levels in a field; periodic in $1/B$, their frequency is proportional to the cross-sectional area of a closed Fermi surface, a direct measure of the carrier density

**Fermi liquid:** a metal that conforms to Landau's Fermi-liquid theory, with signatures that include a coherent (sharp) Fermi surface, the Wiedemann-Franz law, and a $T^2$ dependence of the resistivity

**LSCO:** $La_{2-x}Sr_xCuO_4$

---

## 3. SCATTERING AND PAIRING

The question, then, is this: What makes superconductivity emerge from this particular, rather conventional, metal? The critical doping at which superconductivity springs is roughly the same in all hole-doped cuprates, namely $p_c \approx 0.27$. Note that although it appears to obey weak-coupling BCS theory, at least initially [13, 16], the superconducting state has $d$-wave symmetry [17], rather than the usual $s$-wave symmetry, pointing to an electronic rather than phononic pairing mechanism [18]. What happens at $p_c$ to make $d$-wave pairing prevail? Let us investigate one intriguing clue: At this special doping, the scattering between electrons undergoes a qualitative change. Indeed, it is precisely below $p_c$ that the normal-state electrical resistivity $\rho(T)$ starts to deviate from its quadratic dependence at low temperature [14]. At first, $\rho(T)$ acquires an additional linear term, as in Tl-2201 at $p = 0.25$-$0.26$, where $\rho(T)$ is best described by the form $\rho_0 + AT + BT^2$ below 30 K [12, 13]. At slightly lower doping, $\rho(T)$ becomes purely linear, with $\rho(T) = \rho_0 + AT$ below 80 K or so, as found in $La_{1.6-x}Nd_{0.4}Sr_xCuO_4$ (Nd-LSCO) at $p = 0.24$ [19] and LSCO at $p = 0.23$ [20], both measured down to $T \approx 1$ K in a magnetic field large enough to suppress superconductivity (see **Figure 2**). At still lower doping, the linearity of $\rho(T)$ extends to higher temperature, up to 300 K and above. To describe the broad evolution of $\rho(T)$ with doping, available LSCO data (in zero field) [21] were recently fit to the form $\rho(T) = \rho_0 + AT + BT^2$, over a temperature interval from 200 to 400 K [22]. The result is shown in **Figure 3**, where the parameter $A$ is plotted vs $p$; $A$ is seen to extrapolate to zero at $p = 0.27 = p_c$. In other words, $A \to 0$ at the same doping as $T_c \to 0$. Recent high-field measurements on overdoped LSCO show that the same fit performed over an interval from 1 to 200 K provides a good description of the low-temperature data and leads to the same correlation between $A$ and $T_c$ [20]. Data on Tl-2201 as $p \to p_c$ [12, 14, 23] yield $A \sim T_c$ (see **Figure 4**). This remarkable correlation between linear resistivity and $T_c$ strongly suggests that anomalous (non-Fermi-liquid) scattering and pairing have a common origin. This correlation is supported by ADMR studies of overdoped Tl-2201 which yield an anisotropic linear-$T$ scattering rate that peaks in the same direction as the $d$-wave gap [26] and also scales with $T_c$ [27].

> **Nd-LSCO:** $La_{2-x-y}Nd_ySr_xCuO_4$

Note that the linear resistivity is a universal property of hole-doped cuprates and different materials exhibit the same slope ($A$ coefficient) at a given doping, when measured per $CuO_2$ plane (see [22]). In other words, the anomalous scattering is universal and it switches on at the same doping as superconductivity. The answer to our initial question regarding the cause of $d$-wave pairing would then lie in a second question: What causes the linear temperature dependence of $\rho(T)$ in cuprates? To address this question, we now turn to another family of superconductors, the Bechgaard salts, in which the same correlation between linear resistivity and $T_c$ was recently observed experimentally and elucidated theoretically.

## 4. ORGANIC AND PNICTIDE SUPERCONDUCTORS

The Bechgaard salts $(TMTSF)_2X$ (X = $PF_6$, $ClO_4$) are organic conductors that become superconducting below $T \approx 1$ K [28]. Although they display one-dimensional (1D) conduction at high temperature, their conduction is coherent in two dimensions below approximately 100 K. The phase diagram of $(TMTSF)_2PF_6$ under pressure is shown in **Figure 5** [22, 29]: An SDW phase gives way to a superconducting phase with increasing pressure. The resistivity of $(TMTSF)_2PF_6$, reproduced in **Figure 6**, exhibits the following salient features [22, 29]: a) At a pressure just below the QCP at which SDW order vanishes, $\rho(T)$ undergoes a pronounced upturn at low temperature, upon entering the SDW phase; b) At a pressure just above the QCP, $\rho(T)$ is perfectly linear in temperature down to the lowest temperature; c) At the highest measured pressure, close to where superconductivity disappears, $\rho(T)$ is quadratic in temperature. These are the three regimes characteristic of a quantum phase transition (at $P^*$) [30]: Fermi-surface reconstruction (and gapping) below $P^*$, non-Fermi-liquid (*e.g.* linear-$T$) resistivity at $P^*$, and recovery of the Fermi-liquid $T^2$ dependence beyond $P^*$. Now, the crossover from $T$ to $T^2$ at $T \to 0$ occurs over an interval of pressure that precisely coincides with the interval over which superconductivity exists [22, 29]. In that pressure range, $\rho(T)$ can be fit to the form $\rho_0 + AT + BT^2$, with the $A$ coefficient decreasing monotonically with pressure and scaling with $T_c$ [22, 29], as shown in **Figure 4**. This reveals another instance of the same intimate correlation between linear resistivity and $T_c$ found in the cuprates, and highlighted in the previous section.

> **Organic superconductor:** material made of stacks of organic molecules that conduct with either 1D or 2D character; superconductivity occurs in both versions, with $T_c$ values of order 1 K and 10 K, respectively

The advantage of the Bechgaard salts is that they are well understood theoretically [28]. Weak-coupling renormalization group calculations [31] reproduce the phase diagram – with SDW order giving way to superconductivity – and account in detail for the antiferromagnetic fluctuations observed by nuclear magnetic resonance. This leaves little doubt that pairing and scattering in this material both come from low-energy antiferromagnetic fluctuations. The calculations reveal a fundamental mechanism not considered in previous treatments of non-Fermi-liquid behaviour near an antiferromagnetic QCP: the positive interference between pairing correlations and spin fluctuations [22, 29, 31]. The pairing correlations enhance the spin fluctuations [31] and thereby impart an anomalous linear temperature dependence to the scattering rate [22, 29], causing the resistivity to deviate from the $T^2$ dependence expected at $T \rightarrow 0$ away from the QCP. This interference mechanism operates as long as pairing correlations are significant, and thus provides a natural explanation for the correlation between anomalous (non-Fermi-liquid) resistivity and $T_c$. The fact that the same correlation is present in cuprates is strong evidence that the same positive interference is at play: spin fluctuations cause $d$-wave pairing and $d$-wave correlations enhance scattering. The $d$-wave-like anisotropy of the linear-$T$ scattering detected by ADMR in overdoped Tl-2201 [26] would then be the fingerprint of that interference, the reason why "electrons scatter as they pair" [32].

As non-Fermi-liquid behaviour and unconventional superconductivity are but two manifestations of the same interference, the observation of a non-Fermi-liquid resistivity tied to $T_c$ emerges as a smoking gun for pairing mediated by antiferromagnetic spin fluctuations. It is then interesting to examine other superconductors in this light. The pnictides are a prime testing ground, given that their phase diagram, shown in **Figure 5** for Ba(Fe$_{1-x}$Co$_x$)$_2$As$_2$, is strikingly similar to that of (TMTSF)$_2$PF$_6$, with superconductivity peaking at the QCP where SDW order vanishes. The resistivity of Ba(Fe$_{1-x}$Co$_x$)$_2$As$_2$ [33, 34] shows a linear behaviour near the QCP and a purely quadratic dependence as soon as superconductivity disappears (see **Figure 6**). In the intervening regime, $\rho(T)$ can be fit to the form $\rho_0 + AT + BT^2$, with $A \sim T_c$ [22, 29]. Pnictides therefore provide a third instance of the same correlation between linear resistivity and $T_c$ (see **Figure 4**). This is strong evidence that the same positive interference between antiferromagnetic spin fluctuations and pairing correlations is at play, even though unlike the cuprates and Bechgaard salts the pairing symmetry of pnictides may not be $d$-wave.

The central organizing principle is the presence of a QCP inside the superconducting dome, at which antiferromagnetic/SDW order ends. This is clearly seen in the organic and pnictide superconductors (**Figure 5**), and in several heavy-fermion superconductors (see box) [18, 30, 35, 36]. In the cuprates, the existence, nature and location of such a QCP are all the subject of debate [37, 38]. Below, we discuss several experiments that have recently shed light on the question of a QCP in cuprates.

### Heavy-fermion superconductors

Superconductivity was discovered in $f$-electron "heavy-fermion" materials in 1979 [90], just before its discovery in organic superconductors [91]. Strong evidence that pairing in heavy-fermion metals is of antiferromagnetic origin came from the discovery of superconductivity right at the QCP where antiferromagnetic order vanishes with pressure [18, 35]. A model of magnetic pairing [18, 92] can account, at least qualitatively, for the 10-fold increase in $T_c$ from cubic CeIn$_3$ (0.2 K; [35]) to tetragonal CeRhIn$_5$ (2.3 K; [93]), as a result of the enhanced effectiveness of spin-fluctuation pairing in two dimensions. Several heavy-fermion metals exhibit an antiferromagnetic QCP, and non-Fermi-liquid behaviour is systematically observed in its vicinity [30], with a sub-quadratic temperature dependence of the resistivity. Whether this non-Fermi-liquid behaviour persists away from the QCP, over a range of pressures that coincides with the region of superconductivity, remains to be closely investigated. Existing data on CeRhIn$_5$ [93] suggests that it might. Although the Kondo effect (from the $f$ moments), the strong 3D character and the multi-band Fermi surface all make the problem of scattering and pairing more complex than in the Bechgaard salts, it is not unlikely that antiferromagnetic spin fluctuations play fundamentally the same role and the quantum-critical behaviour is again modified by the positive interference of pairing correlations.

# 5. QUANTUM CRITICAL POINT AND BROKEN TRANSLATIONAL SYMMETRY

**Figure 6** shows that Nd-LSCO exhibits the same three regimes of quantum criticality that are seen in the organic and pnictide superconductors, where they are associated with a SDW QCP. The upturn in $\rho(T)$ at $p = 0.20$ below 40 K (**Figures 2** and **6**) is caused by a reconstruction of the Fermi surface, as confirmed by a parallel upturn in the Hall [19] and Seebeck [39] coefficients. While $R_H(T\to 0)$ is consistent with a single large hole-like Fermi surface at $p = 0.24$, being equal to $1/e(1+p)$, it suddenly becomes much larger at $p = 0.20$, and then much smaller (almost negative) at $p = 0.12$ [40]. A similar evolution is observed in the Seebeck coefficient $S$, whereby $S/T$ at $T \to 0$ goes from small positive at $p = 0.24$ to large positive at $p = 0.20$ [39], and then to large negative at $p = 0.12$ [40, 41]. These dramatic changes in $R_H$ and $S$ are typical of a Fermi surface whose topology changes with doping and includes both hole-like and electron-like portions. In Nd-LSCO and the closely related Eu-doped LSCO (Eu-LSCO), the mechanism responsible for the reconstruction of the large Fermi surface is clear [42]: it is the onset of "stripe" order, a form of SDW order first detected by neutron diffraction [43], with an associated charge-density-wave order [43], also detected by X-ray diffraction [44] and nuclear quadrupole resonance [45]. **Figure 7** shows how in Eu-LSCO at $p = 1/8$ the onset of stripe order coincides with the drop in $R_H(T)$ [46], which also coincides with the drop in $S/T$ [41]. As shown in the phase diagram of **Figure 8**, the onset of stripe order in Nd/Eu-LSCO occurs at a temperature $T_{CO}$, which peaks at $p = 1/8$ and decreases monotonically with doping, extrapolating to zero at $p \approx 0.24$. The stripe phase is most stable at 1/8 because at that doping its period is commensurate with that of the lattice [25, 45].

> **Eu-LSCO:** $La_{2-x-y}Eu_ySr_xCuO_4$
>
> **Stripe order:** unidirectional spin or charge density-wave order; the modulation is in general incommensurate with the crystal lattice, and it breaks translational and rotational symmetries; while both spin and charge stripes are seen in Nd-LSCO and Eu-LSCO, only spin stripes have so far been seen in YBCO and LSCO

Therefore, the QCP in Nd-LSCO marks the onset of a finite-$Q$ modulation of the spin and charge densities at $T = 0$ which breaks the translational symmetry of the lattice and hence causes a reconstruction of the Fermi surface. (It is possible that spin and charge order set in at somewhat different dopings and temperatures.) The critical doping at which this symmetry-breaking and Fermi-surface reconstruction onset was pinpointed by tracking the upturn in the $c$-axis resistivity of Nd-LSCO [47], giving $p^* = 0.235 \pm 0.005$. A similar study performed on $Bi_2Sr_2CaCu_2O_{8+\delta}$ gave a comparable value of $p^*$ [48]. Calculations [49] show that stripe order does cause the Fermi surface to break up into small electron and hole pockets (plus some quasi-1D sheets), and these can give rise to positive and negative swings in $R_H(T\to 0)$ as the SDW potential grows with underdoping [50].

This establishes the existence of a QCP inside the superconducting dome, at which stripe order (a form of SDW order) ends, much as in the organic and pnictide superconductors. The analogy then suggests that fluctuations of the stripe order are responsible for the linear resistivity and, given the correlation with $T_c$, the pairing. In support of this connection, the strength of antiferromagnetic fluctuations in overdoped LSCO measured by inelastic neutron scattering has been shown to scale with $T_c$ [51]. Two important questions now arise: Is stripe order a generic property of hole-doped cuprates? Is the pseudogap phase related to stripe order? We address the first question in the remainder of this section, and explore the second question in section 6.

Scanning tunneling microscopy (STM) studies have revealed real-space modulations of the charge density in three different hole-doped cuprates [52, 53, 54, 80, 94]: $Bi_2Sr_2CaCu_2O_{8+\delta}$, $Ca_{2-x}Na_xCuO_2Cl_2$ and $Bi_2Sr_2CuO_{6+x}$. These have stripe-like unidirectional character on the nanometer scale [55, 80]. The modulations persist into the overdoped regime and their real-space period appears to lengthen with doping [54], which points to a charge-density-wave order driven by Fermi-surface nesting. Neutron scattering studies have revealed stripe-like SDW order in LSCO for dopings below $p_S \approx 1/8$ [56]. This critical doping moves up in a magnetic field [56, 57], such that the QCP is expected to be roughly at $p^* \approx 0.2$ once superconductivity has been fully suppressed. The fact that the QCP moves up with field, from $p_S$ in the superconducting state up to $p^*$ in the normal state (**Figure 1**) is attributed to a competition between SDW and superconducting phases [58, 59, 95]. In Nd-LSCO, where stripe order is stronger, the presence of a weakened superconductivity has little effect on $p^*$, and hence $p_S \approx p^* = 0.235$. In LSCO, superconductivity is stronger and its presence does shift the QCP down significantly. Taken together, STM, neutron and X-ray studies on several different materials make a strong case that stripe order is a generic tendency of hole-doped cuprates at low temperature (for reviews on stripe order and fluctuations, see [60, 96]).

> **STM:** scanning tunnelling microscopy
>
> **YBCO:** $YBa_2Cu_3O_y$

Because of its high maximal $T_c$ and low level of disorder, the case of $YBa_2Cu_3O_y$ (YBCO) deserves special attention; any phenomenon deemed generic should be seen in YBCO. Muon spin relaxation has shown that magnetism is present in YBCO below $p \approx 0.08$ [61], and neutron diffraction has revealed spin-stripe SDW order in YBCO, but again only up to $p \approx 0.08$ [62]. Although it is quite conceivable that in zero field the SDW phase is confined to such low doping because of a particularly strong competition from superconductivity (see [63]), it is important to establish whether SDW order persists up to higher doping in the absence of such competition. A number of recent studies in high magnetic fields provide compelling evidence that it does.

The observation of quantum oscillations in YBCO at $p = 0.10$-$0.11$ [64, 65] revealed the existence of a small closed pocket in the Fermi surface of an underdoped cuprate at $T \to 0$ (see **Figure 9**), whose k-space area is some 30 times smaller than the area enclosed by the large hole-like cylinder characteristic of the overdoped regime [11]. Similar oscillations were also observed in $YBa_2Cu_4O_8$, for which $p \approx 0.14$ [66, 67]. The fact that the Hall coefficient of both materials is large and negative at $T \to 0$ (see **Figure 10**) indicates that this small closed Fermi pocket is in fact electron-like [68]. The normal-state Seebeck coefficient reaches a negative value of $S/T$ as $T \to 0$ which is quantitatively consistent with the frequency and cyclotron mass of the quantum oscillations only if those come from an electron Fermi pocket [41]. The very existence of an electron pocket in a hole-doped cuprate is compelling evidence of broken translational symmetry, the result of a Fermi-surface reconstruction caused by the onset of some new periodicity [69]. In YBCO at $p = 0.12$, $R_H(T)$ starts its descent to negative values upon cooling in precisely the same way as it does in Eu-LSCO at $p = 0.125$ [42], where this drop is associated unambiguously with the onset of stripe order (see **Figure 7**). The same striking similarity between YBCO and Eu-LSCO is observed in the way that $S/T$ falls to negative values [41], pointing again to the same underlying mechanism, the onset of stripe order. In YBCO, this mechanism is still present at $p \approx 0.14$ and extrapolation suggests that $p^* > 0.2$. Taken together, these high-field data support the case that the normal-state QCP identified in Nd-LSCO at $p^* = 0.235$ is also present in YBCO and is therefore a generic property of hole-doped cuprates, once the competing superconducting phase has been removed.

Note that the temperature below which Fermi-surface reconstruction in YBCO begins (*i.e.* where $R_H$ and $S/T$ start to fall) is maximal at $p = 1/8$ [68], the doping where $T_c$ is most strongly suppressed relative to its ideal parabolic dependence on doping [70]. The fact that peak (in $R_H$ maximum) and dip (in $T_c$) coincide is consistent with a scenario of competing stripe and superconducting orders, the former being stabilized by commensurate locking with the lattice at $p = 1/8$, as in the $La_2CuO_4$-based cuprates. Note also that the electron-pocket state is not induced by the magnetic field: at $p = 1/8$, the drop in $R_H(T)$ is observed in the limit of zero field and is independent of field [68, 71]. The field simply serves to remove superconductivity and allow transport measurements to be extended to the $T \to 0$ limit.

## 6. PSEUDOGAP PHASE AND BROKEN ROTATIONAL SYMMETRY

Above we focused on the $T \to 0$ limit and argued that there is a generic normal-state QCP in the overdoped regime of hole-doped cuprates below which translational symmetry is broken and the large hole-like Fermi surface is reconstructed. Now we examine this same process as a function of temperature. In other words, after having investigated a *p*-cut at $T = 0$ in the phase diagram (**Figure 1**), across the QCP at $p^*$, we now look at a *T*-cut at $p < p^*$, across the pseudogap temperature $T^*$.

We begin by defining $T^*$ as the temperature $T_\rho$ below which the in-plane resistivity $\rho(T)$ deviates from its linear temperature dependence at high temperature – a standard definition of $T^*$ in YBCO [8, 72]. In Nd-LSCO, $T_\rho$ marks the onset of an upward deviation in $\rho(T)$, which eventually leads to an upturn at low temperature (**Figure 2**). In **Figure 8**, $T_\rho$ is plotted as a function of doping; it goes to zero at $p^*$. Note that $T_{CO}$, the onset of long-range stripe order, which also vanishes roughly at $p^*$, lies well below $T_\rho$, so that $T_\rho \approx 2\ T_{CO}$. This suggests that $T_\rho$ marks the onset of stripe fluctuations and that the pseudogap phase below $T^*$ is initially just a short-range / fluctuating precursor of the order that eventually develops fully at lower temperature [42, 73].

As a probe of electronic transformations and phase transitions, the Nernst effect is in general vastly more sensitive than resistivity [74], in essence because changes in carrier density and scattering rate tend to combine in the former whereas they tend to cancel in the latter. Nernst measurements have been used only recently to study the onset of the pseudogap phase in cuprates [46, 75, 76]. A pseudogap temperature $T_\nu$ can be defined from the Nernst coefficient $\nu(T)$ in much the same way as for $\rho(T)$, namely as the temperature below which $\nu/T$ deviates from its linear temperature dependence at high temperature [46, 76, 77]. The resulting phase diagram is shown in **Figure 11a** for LSCO, Eu-LSCO and Nd-LSCO and in **Figure 11b** for YBCO. First, we see that $T_\nu = T_\rho$, within error bars. This shows that both $\rho$ and $\nu$ detect the same pseudogap temperature $T^*$, which is not surprising since $\nu$ involves the energy derivative of the conductivity [74]. Second, $T_\nu$ is the same in LSCO and Eu/Nd-LSCO. This shows that the onset of the pseudogap phase is independent of the detailed crystal structure and the relative strength of stripe order and superconductivity. It also strongly suggests that the elusive normal-state QCP in LSCO [20] is located at the same doping $p^*$ as it is in Nd-LSCO (namely at $p \approx 0.24$), or close to it. Thirdly, $T_\nu$ in YBCO can be tracked all the way up to $p = 0.18$ [76], the highest doping achievable in pure YBCO (at full oxygen content [70]). Comparison with the LSCO phase diagram suggests that $T_\nu$ in YBCO will extrapolate to zero at much the same $p^*$. This is further support for a generic normal-state QCP in hole-doped cuprates at $p^* \approx 0.24$.

**Nernst effect:** transverse electric field $E_y$ across the width of a metallic sample that develops when a temperature gradient $\partial T / \partial x$ is applied along its length in the presence of a perpendicular magnetic field $B$; the Nernst coefficient is defined as $\nu = E_y / (B\ \partial T / \partial x)$

The suggestion that $T^*$ marks the onset of stripe fluctuations (or short-range order) has recently received strong support from a study of the Nernst effect in untwinned crystals of YBCO [76]. Measurements with the temperature gradient applied along the *a* axis and then the *b* axis of the orthorhombic lattice reveal a pronounced anisotropy that grows with decreasing temperature starting precisely at $T^*$ throughout the phase diagram and reaching values as high as $\nu_b / \nu_a = 7$ before superconductivity intervenes [76] (see **Figure 12**). These findings are consistent with prior evidence of in-plane anisotropy in the resistivity [78] and in the spin fluctuation spectrum [62], detected below $p \approx 0.08$. The Nernst data now provide the missing link to the pseudogap phase by showing that $T^*$ marks the onset of broken rotational symmetry in the electronic properties of the $CuO_2$ planes. This unidirectional character is one of the defining signatures of stripe order [60, 73, 96]. Microwave and STM studies have provided complementary evidence of broken rotational symmetry, observed at low temperature in the superconducting state. The microwave conductivity of YBCO exhibits a strong in-plane anisotropy at $p = 0.1$ which is not present at $p = 0.18$ [79], suggesting that the zero-field QCP in YBCO lies between those two dopings, *i.e.* $0.1 < p_S < 0.18$ (see **Figure 1**). STM revealed that rotational symmetry is broken on the local scale at the surface of two cuprates [55, 80], in the simultaneous presence of broken translational symmetry [55]. This glassy nanostripe order was recently linked to the pseudogap energy scale [81].

In summary, the following picture of the pseudogap phase is emerging:

1) All hole-doped cuprates have a similar $T^*$ line that ends at a universal critical point $p^*$, near 0.24 in the absence of superconductivity.

2) Below this normal-state QCP, the large hole-like Fermi surface characteristic of the overdoped regime is reconstructed into several pieces, including electron-like pockets and hole-like sheets.

3) This reconstruction is caused by the onset of stripe order, which breaks the translational symmetry of the lattice at low temperature.

4) With increasing temperature, the stripe-ordered phase ends well before the pseudogap crossover temperature $T^*$; the intervening region of the phase diagram is most likely a regime of fluctuating short-range stripe order, which breaks the four-fold rotational symmetry of the lattice.

5) Just to the right of the $T^*$ line, and down to $T = 0$ at $p^*$, the resistivity is linear in temperature; the scattering mechanism responsible for this non-Fermi-liquid behavior is most likely the fluctuations of the pseudogap phase, *i.e.* stripe fluctuations.

6) The intimate connection between linear resistivity and $T_c$ strongly suggests that scattering and pairing have a common origin, rooted in the fluctuations of the ordered phase, a type of SDW order.

## 7. ELECTRON-DOPED CUPRATES

Above we only considered hole-doped cuprates. Now we turn to the electron-doped cuprates; for a recent review of electron-doped cuprates, see [82]. The picture that has emerged over the past few years on that side of the phase diagram is remarkably similar to what is summarized in the last section. This section lists the key findings, obtained mostly from two materials, $Pr_{2-x}Ce_xCuO_4$ and $Nd_{2-x}Ce_xCuO_4$ :

1) The pseudogap crossover temperature $T^*$ decreases with $x$ and vanishes at a critical electron doping $x^* \approx 0.17$ [82].

2) From Hall effect data in the $T \rightarrow 0$ limit [83], there is a normal-state QCP at $x^* \approx 0.165 \pm 0.005$, below which the large hole-like Fermi surface is reconstructed into small electron and hole pockets.

This reconstructed topology is confirmed by ARPES [84], with which both electron and hole pockets have been seen. The transition from small to large Fermi surface was recently detected via quantum oscillation measurements across $x^*$ [85].

3) The observed reconstruction can be accounted for in a model of commensurate SDW (antiferromagnetic) order, which breaks the translational symmetry of the lattice below $x^*$ [86].

4) Long-range antiferromagnetic order has been observed by neutron scattering to set in below a temperature $T_N$ that appears to vanish at a critical doping somewhat lower than $x^*$, namely at $x \approx 0.13$ [87]. The separation between this zero-field onset of antiferromagnetic order and the in-field normal-state QCP at $x^*$ may again result from the competing effect of superconductivity [59].

5) $T^*$ lies significantly above $T_N$, with $T^* \approx 2\, T_N$; because $T^*$ is roughly the temperature below which the antiferromagnetic correlation length exceeds the thermal de Broglie wavelength of the charge carriers [87], the pseudogap phase has been interpreted [88] as a regime of short-range antiferromagnetic correlations (the so-called "renormalized classical regime").

6) The resistivity exhibits the three regimes of quantum criticality [83]: perfectly linear in temperature down to the lowest temperature at $x^*$ [89]; upturn at low temperature for $x < x^*$; curvature, approaching $T^2$, for $x > x^*$.

The same basic scenario invoked for the hole-doped side applies here on the electron-doped side, with a QCP as the central organizing principle. Below that QCP, there is SDW order, broken translational symmetry and Fermi-surface reconstruction. Coming down in temperature, a two-stage evolution occurs, where short-range SDW correlations / fluctuations appear first below $T^*$ and long-range order only sets in later. The fact that the SDW order is commensurate, with wavevector $Q = (\pi, \pi)$, means that the four-fold rotational symmetry of the lattice is not broken in this case. At the QCP, where $T_c$ peaks (in zero field), the resistivity is perfectly linear, almost certainly the result of scattering by antiferromagnetic fluctuations. (A linear resistivity as $T \to 0$ has only ever been observed on the border of antiferromagnetic order.) There is only one missing piece to complete the mirror-like symmetry of the phenomena on both sides of the cuprate phase diagram: a correlation between linear resistivity and $T_c$ in the electron-doped materials. I predict that this correlation will be found there too.

## 8. CONCLUSION

Stepping back to look at the high-$T_c$ puzzle from a distance, incorporating both sides of the cuprate phase diagram and bearing in mind the broad landscape of unconventional superconductivity as we have come to know it over the past 30 years, the essential ingredient appears to be a QCP at which SDW order comes to an end. Associated with this QCP are antiferromagnetic spin fluctuations that can mediate anisotropic pairing and scatter electrons in a way which is in turn strongly influenced by the pairing correlations. The intimate connection between pairing and scattering observed in organic, pnictide and cuprate superconductors is the tell-tale sign that antiferromagnetism and superconductivity work hand in hand in those materials. This positive interference between spin fluctuations and pairing correlations is a new avenue to be explored in our quest for a room-temperature superconductor.

**SUMMARY POINTS**

1. There is a QCP in both hole-doped and electron-doped cuprates at which the large hole-like Fermi surface of the overdoped regime is reconstructed.

2. The reconstructed Fermi surface is typically made of both small electron pockets and other hole-like surfaces. This reconstruction affects all electronic properties of these materials, including their ability to form a superconducting state.

3. The reconstruction is caused by the spontaneous onset of a new spatial periodicity that breaks the translational symmetry of the lattice. The order is most likely commensurate antiferromagnetism on the electron-doped side and stripe order – a form of unidirectional and generally incommensurate spin-density-wave – on the hole-doped side. Stripe order also breaks the rotational symmetry of the tetragonal $CuO_2$ planes.

4. The pseudogap temperature $T^*$ marks the onset of antiferromagnetic correlations in electron-doped cuprates and the onset of a strong in-plane anisotropy in the transport properties of hole-doped cuprates, most likely due to anisotropic spin fluctuations.

5. The QCP is the end of the pseudogap boundary, where the crossover temperature $T^*$ goes to zero. It is located inside the region of superconductivity in the phase diagram.

6. Right at the QCP, in both electron- and hole-doped cuprates, the resistivity is perfectly linear in temperature as $T \to 0$. Only ever observed on the border of SDW order, such a linear resistivity is attributed to the scattering of charge carriers by antiferromagnetic spin fluctuations.

7. A linear resistivity as $T \to 0$ was recently observed in both organic and pnictide superconductors, again on the border of SDW order.

8. The strength of this linear resistivity is found to scale with $T_c$ in cuprate, organic and pnictide superconductors. This direct empirical correlation strongly suggests that pairing and linear-$T$ scattering have a common origin, most likely antiferromagnetic spin fluctuations in all three families of materials.

**FUTURE ISSUES**

1. Why is the $T_c$ of cuprates higher on the hole-doped side ?

2. Why does the electron system in cuprates have a preference for unidirectional incommensurate order on the hole-doped side ?

3. Is charge order in hole-doped cuprates fundamental or simply secondary to spin order ?

4. How does the underdoped metal, with a Fermi surface that is still coherent at $p = 0.1$, turn into a Mott insulator as $p \to 0$ ?

5. Is the positive interference between pairing correlations and spin fluctuations, discovered in studies of organic superconductors, a general mechanism for non-Fermi-liquid behaviour near an antiferromagnetic QCP ?

## ACKNOWLEDGMENTS

I would like to thank Kamran Behnia, Claude Bourbonnais, Andrey Chubukov, Nicolas Doiron-Leyraud, Richard Greene, Catherine Kallin, Steve Kivelson, Gilbert Lonzarich, Andrew Millis, Michael Norman, Cyril Proust, Subir Sachdev, André-Marie Tremblay, and Matthias Vojta for helpful discussions. I would like to acknowledge the support of the Canadian Institute for Advanced Research and funding from NSERC, FQRNT, CFI, and a Canada Research Chair.

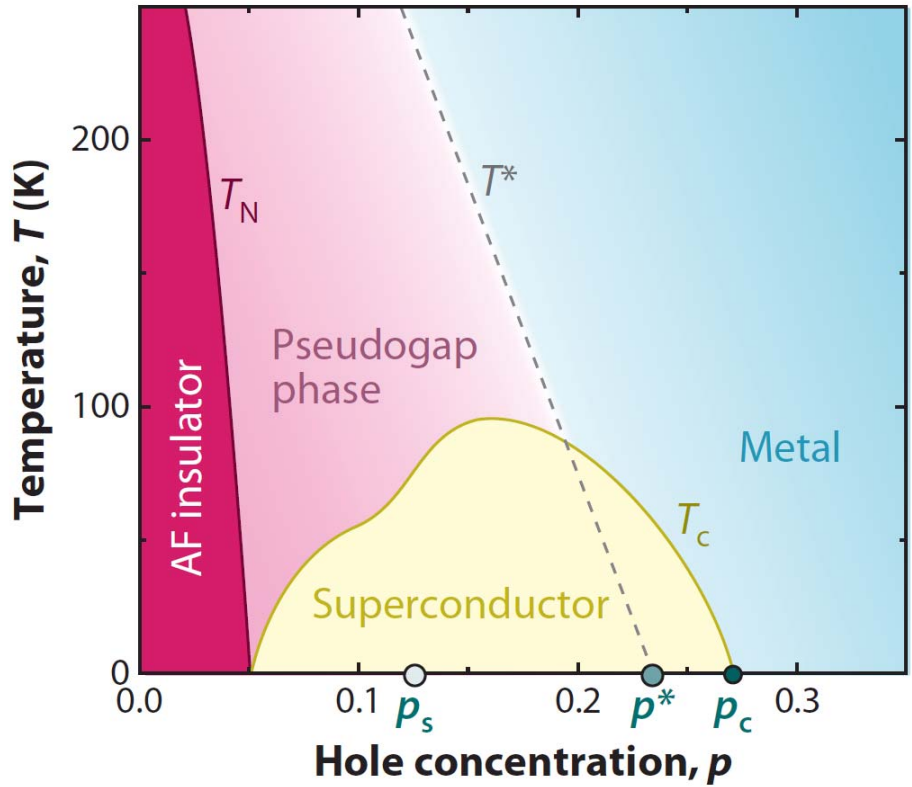

**Figure 1. Phase diagram of cuprate superconductors.**

Schematic phase diagram of cuprate superconductors as a function of hole concentration (doping) $p$. The Mott insulator at $p = 0$ shows antiferromagnetic (AF) order below $T_N$, which vanishes rapidly with doping. At high doping, the metallic state shows all the signs of a conventional Fermi liquid. At the critical doping $p_c$, two events happen simultaneously: superconductivity appears (below a critical temperature $T_c$) and the resistivity deviates from its Fermi-liquid behaviour, acquiring a linear temperature dependence. The simultaneous onset of $T_c$ and linear resistivity is the starting point for our exploration of cuprates. The evolution from metal to insulator is interrupted by the onset of the "pseudogap phase" which sets in below a crossover temperature $T^*$, which goes to zero at a quantum critical point (QCP) located at $p^*$ in the absence of superconductivity (removed for example by application of a large magnetic field). The existence, nature and location of such a QCP are a major focus of this review. In the presence of superconductivity, the QCP may move to lower doping, down to $p_S$, as a result of a competition between the pseudogap and superconducting phases [59, 95].

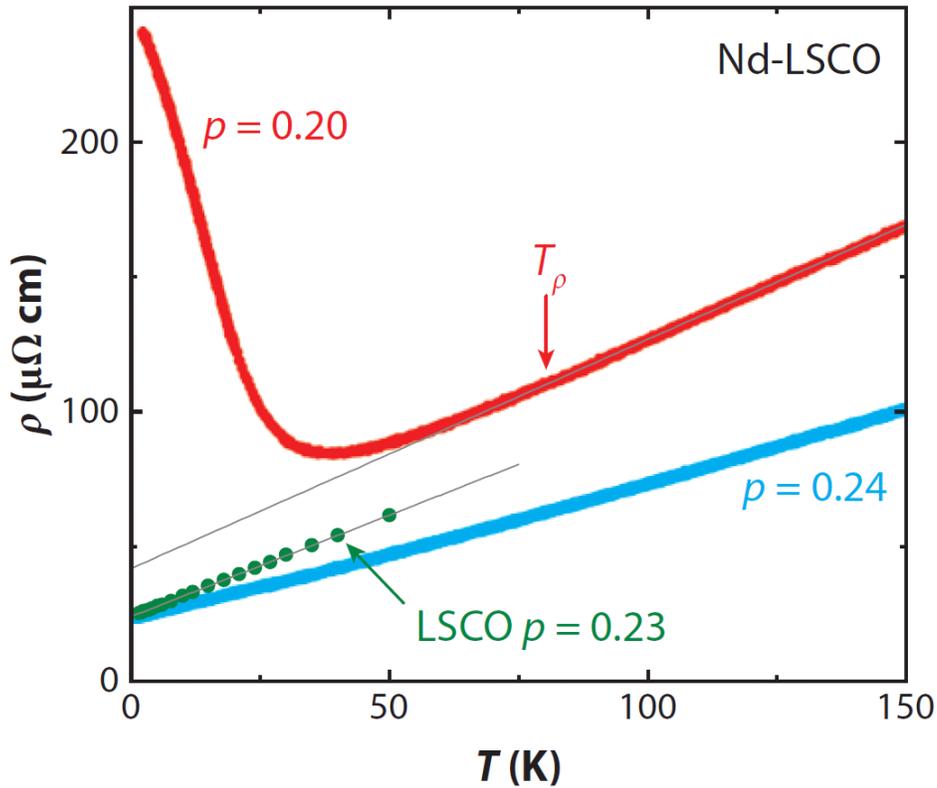

**Figure 2. Electrical resistivity of overdoped cuprates.**

In-plane electrical resistivity of two hole-doped cuprates once superconductivity has been suppressed by application of a sufficiently large magnetic field $B$: Nd-LSCO at $p = 0.24$ ($B = 33$ T; blue) and at $p = 0.20$ ($B = 35$ T; red) (data from [19]); LSCO at $p = 0.23$ ($B = 45$ T; green) (data from [20]). The data for Nd-LSCO at $p = 0.24$ and LSCO at $p = 0.23$ are both perfectly linear at low temperature, down to at least 1 K. At lower doping, the resistivity of both materials deviates upward from linearity below a certain temperature. This effect is more pronounced in Nd-LSCO, where it leads to a large upturn, as shown here for $p = 0.20$. $T_\rho$ is the temperature below which the resistivity begins to deviate from its linear dependence at high temperature. Grey lines are linear fits.

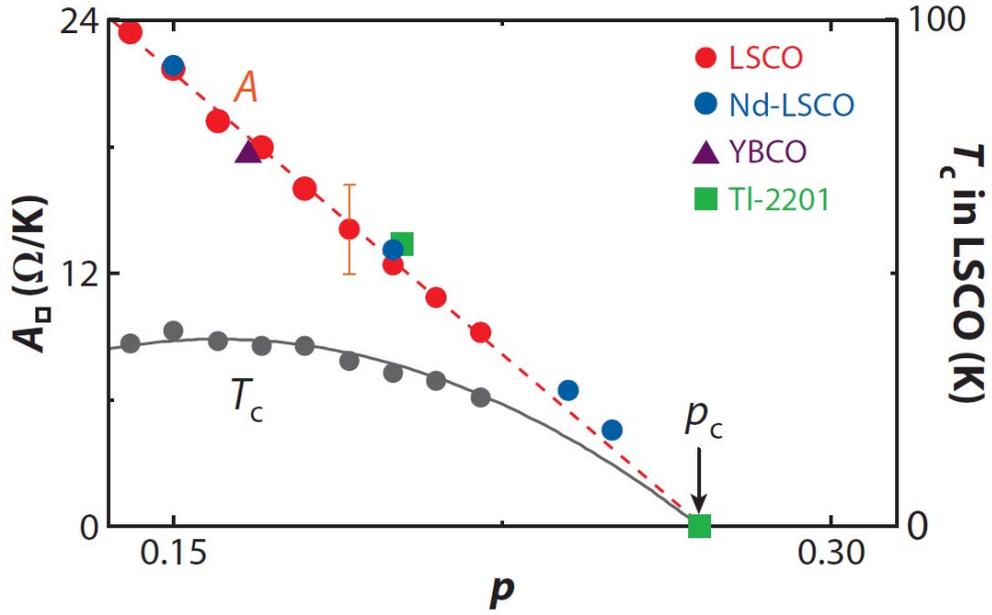

**Figure 3. Linear resistivity and $T_c$ vs doping in hole-doped cuprates.**

Coefficient of the linear resistivity of cuprates per $CuO_2$ plane, $A_\square = A/d$, as a function of doping $p$, for LSCO (red dots; [21, 24]), Nd-LSCO (blue dots; [19, 25]), YBCO (purple triangle; [21]), and Tl-2201 (green squares; [14, 23]). The data are extracted from fits [22] of the form $\rho_0 + AT + BT^2$ to published data. The red dashed line is a linear fit to the LSCO data points. The grey dots are the corresponding $T_c$ for LSCO [21]. The grey line is the formula $T_c = T_c^{max} [1 - 82.6 (p - 0.16)^2]$, with $T_c^{max} = 37$ K. Note that the coefficient of the linear term goes to zero where superconductivity vanishes, *i.e.* $A \to 0$ as $T_c \to 0$, at $p_c = 0.27$. Figure adapted from [22].

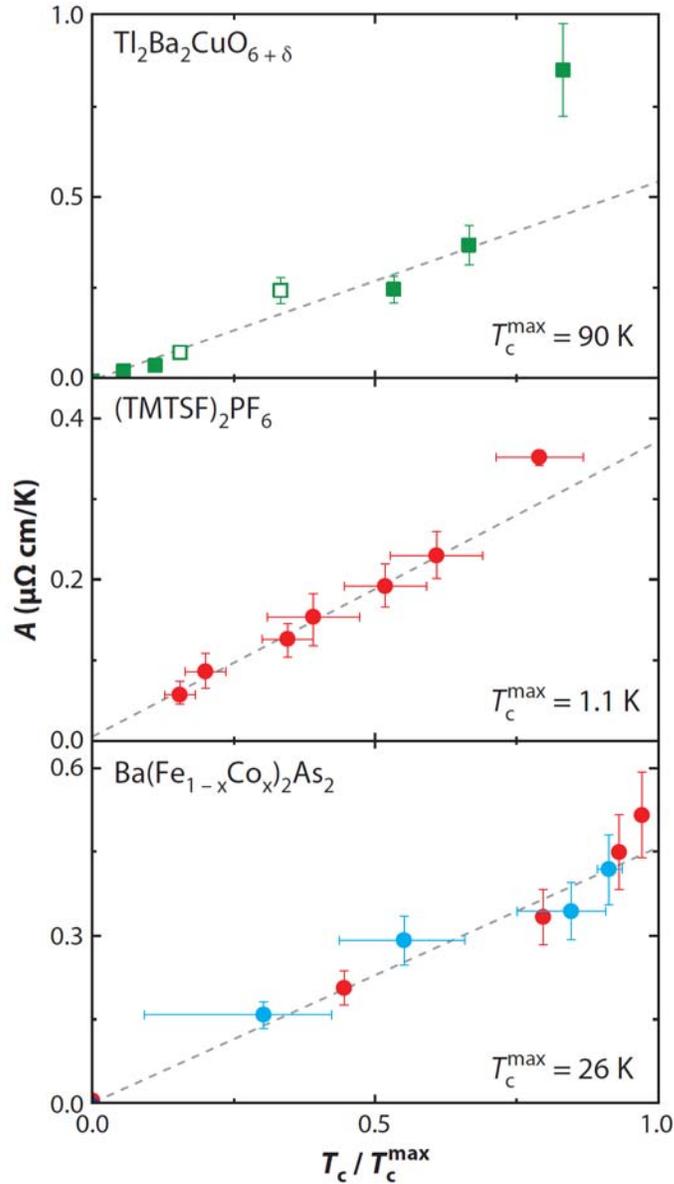

**Figure 4. Correlation between linear resistivity and $T_c$ in cuprate, organic and pnictide superconductors.**

Coefficient of the linear term $A$ in the resistivity $\rho(T)$ as a function of $T_c$ for the cuprate Tl-2201 (top; closed squares, data from [14]; open squares, data from [12, 23]), the Bechgaard salt $(TMTSF)_2PF_6$ (middle; data from [22, 29]) and the pnictide $Ba(Fe_{1-x}Co_x)_2As_2$ (bottom; red dots, data from [34]; blue dots, data from [33]). The $T_c$ values correspond to different hole dopings, pressures and cobalt concentrations, respectively, plotted normalized to $T_c^{max}$, the maximum value of $T_c$ in the phase diagram (as indicated). The $A$ coefficient is obtained from fits of $\rho(T)$ to the form $\rho_0 + AT + BT^2$ (see [22] for top and [29] for middle and bottom panels). Adapted from [22] and [29].

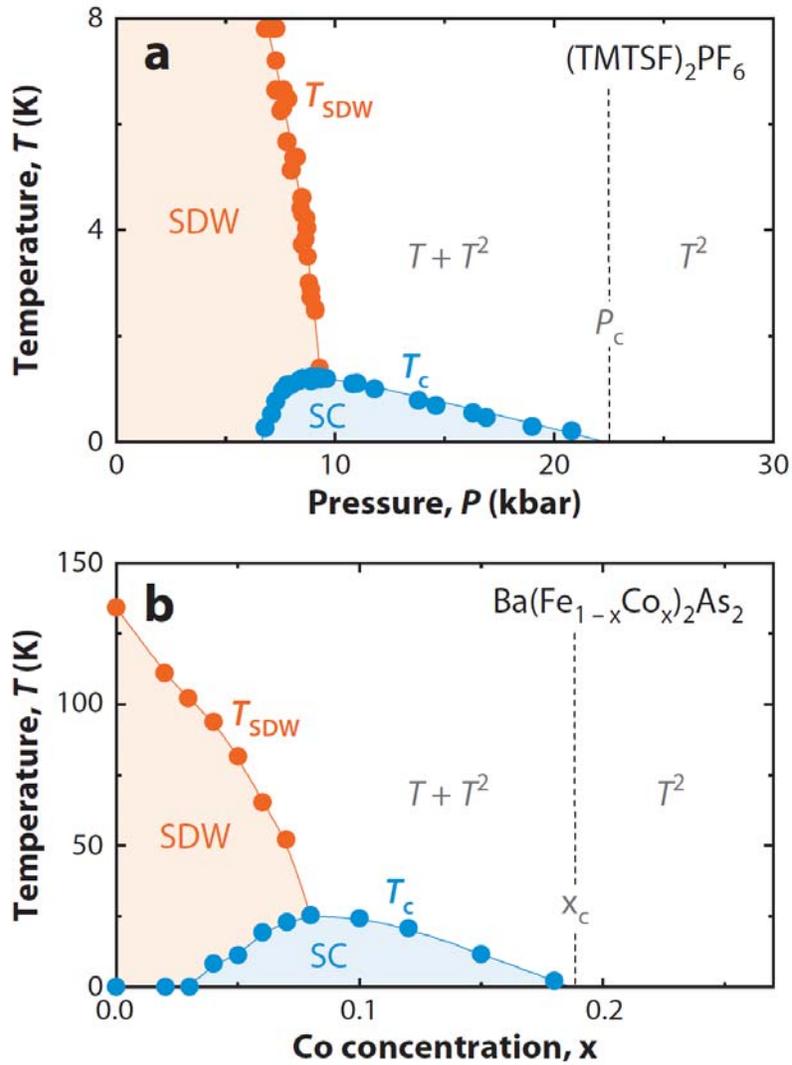

**Figure 5. Phase diagram of organic and pnictide superconductors.**

**a)** Temperature-pressure phase diagram of $(TMTSF)_2PF_6$, showing a spin-density-wave (SDW) phase below $T_{SDW}$ (orange dots) and superconductivity (SC) below $T_c$ (blue dots) [22, 29]. The latter phase ends at the critical pressure $P_c$. **b)** Temperature-doping phase diagram of the iron-pnictide superconductor $Ba(Fe_{1-x}Co_x)_2As_2$, as a function of nominal Co concentration $x$, showing a metallic SDW phase below $T_{SDW}$ and superconductivity below a $T_c$ which ends at the critical doping $x_c$ [34]. In both panels the vertical dashed line separates a regime where the resistivity $\rho(T)$ grows as $T^2$ (on the right) from a regime where it grows as $T + T^2$ (on the left). From [29].

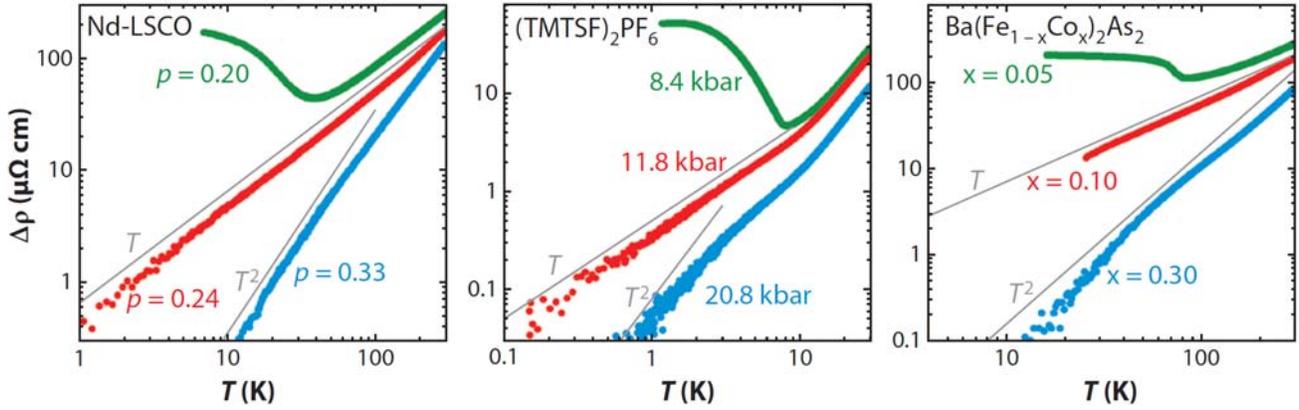

**Figure 6. Quantum criticality in the resistivity of cuprate, organic and pnictide superconductors.**

Temperature dependent part of the in-plane normal-state resistivity of materials in three families of superconductors, plotted as $\rho(T) - \rho_0$ vs $T$ on a log-log scale. Three values of the relevant tuning parameter were chosen: below, at and above their respective quantum critical points (QCPs). Left panel (from [39]): data on hole-doped cuprates Nd-LSCO at $p = 0.20$ and $p = 0.24$ (from [19]) and LSCO at $p = 0.33$ (from [15]); the QCP at a hole doping $p^* \approx 0.24$ marks the end of the stripe-ordered phase in Nd-LSCO [19, 25]. Middle panel (from [29]): data on the organic Bechgaard salt $(TMTSF)_2PF_6$; the QCP at a pressure $P^* \approx 10$ kbar marks the end of the SDW phase. Right panel: data on the pnictide $Ba(Fe_{1-x}Co_x)_2As_2$ (from [34]); the QCP at a Co concentration $x^* \approx 0.10$ marks the end of the SDW phase. Note in all cases: a linear dependence as $T \to 0$ at the QCP; a Fermi-liquid $T^2$ dependence above the QCP (beyond the superconducting phase); an upturn caused by Fermi-surface reconstruction upon entry into the ordered phase below the QCP.

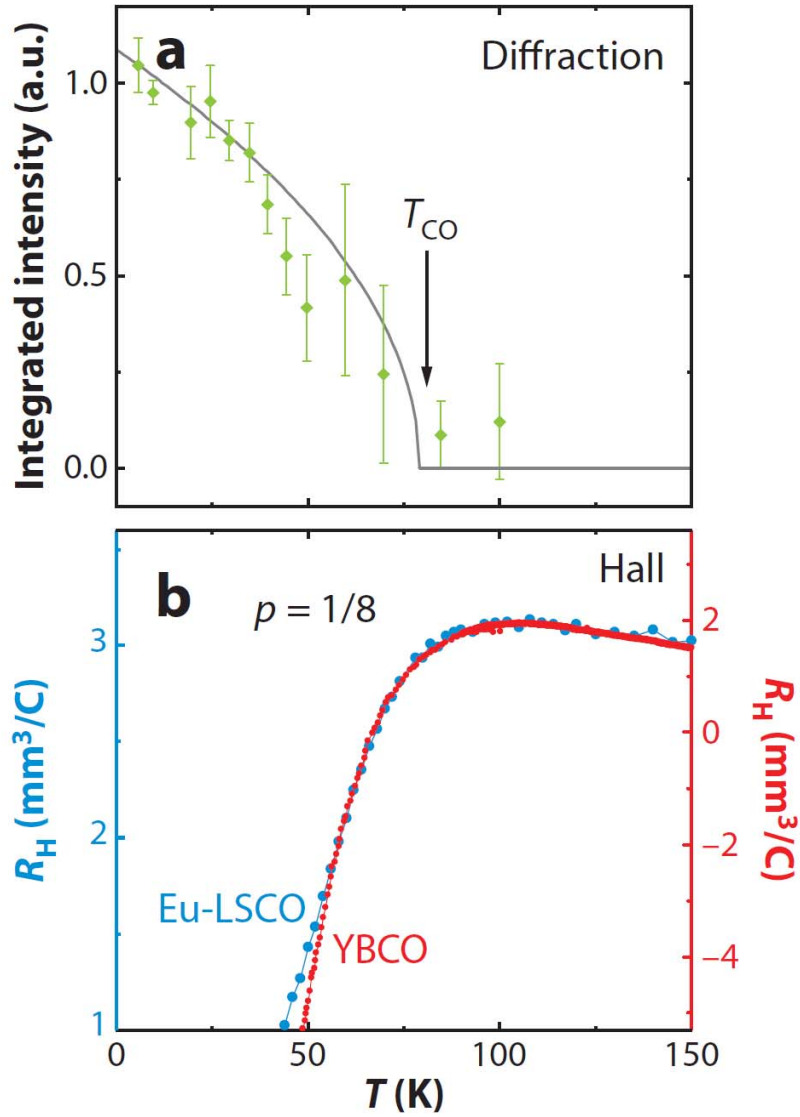

**Figure 7. Stripe order and Hall coefficient in cuprates at $p = 1/8$.**

**a)** Temperature dependence of charge stripe order in Eu-LSCO at $p = 1/8$, as detected by resonant soft X-ray diffraction (data from [99]). The grey line is a guide to the eye. **b)** Hall coefficient vs temperature measured in $B = 15$ T for Eu-LSCO (blue, left axis; from [46]) and YBCO (red, right axis; from [68]), both at $p \approx 1/8$. Adapted from [42].

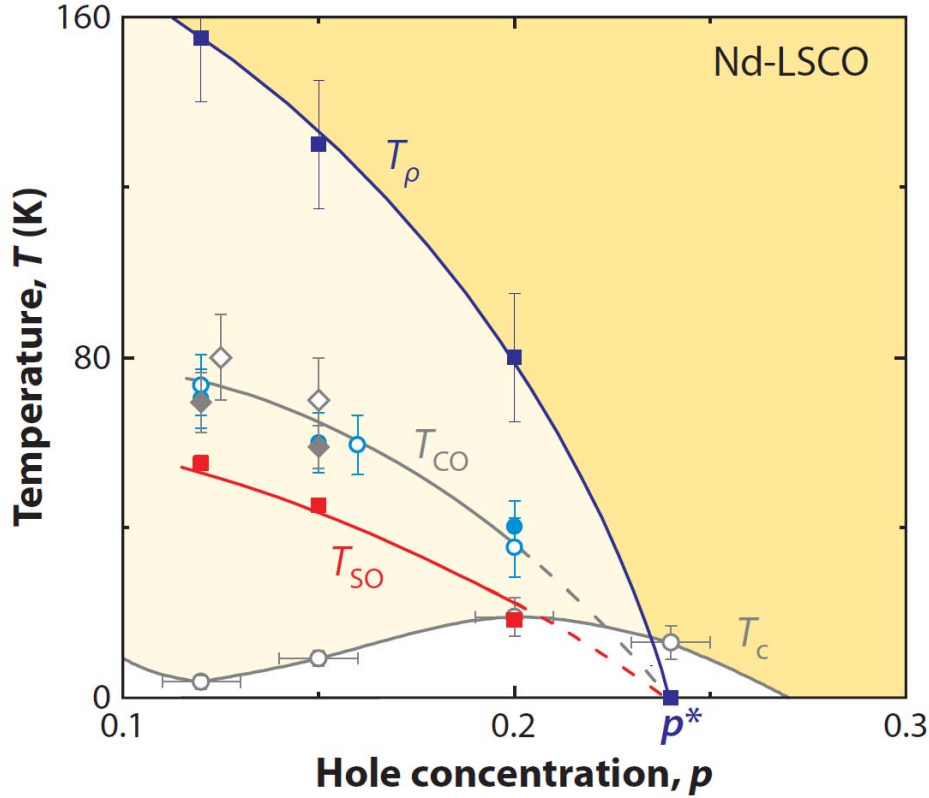

**Figure 8. Phase diagram of Nd-LSCO.**

Temperature-doping phase diagram of Nd-LSCO showing the superconducting phase below $T_c$ (open grey circles) and the pseudogap region delineated by the crossover temperature $T_\rho$ (dark blue squares). Also shown is the region where magnetic order is observed below $T_{SO}$ (red squares) and charge order is detected below $T_{CO}$ (grey diamonds and blue circles). These onset temperatures are respectively defined as the temperature below which: the resistance is zero [19, 25]; the in-plane resistivity $\rho(T)$ deviates from its linear dependence at high temperature [19, 25]; magnetic Bragg peaks are observed in neutron diffraction [25]; charge order is detected by either X-ray diffraction (on Nd-LSCO, closed grey diamonds, from [100]; on Eu-LSCO, open grey diamonds, from [99]) or nuclear quadrupole resonance (data from [45]: Nd-LSCO, closed blue circles; Eu-LSCO, open blue circles). The central feature is the QCP at a doping $p^*$ inside the superconducting region. Adapted from [42].

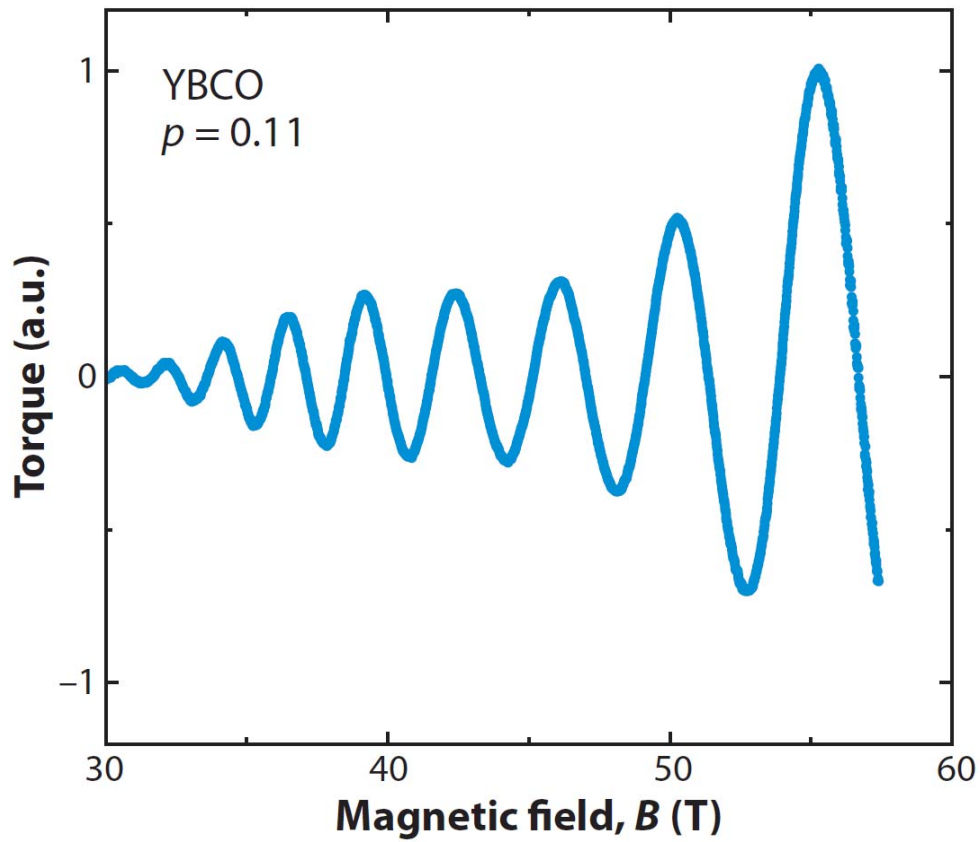

**Figure 9. Quantum oscillations in YBCO.**

Quantum oscillations in the magnetization of YBCO at $p = 0.11$, detected by torque magnetometry as a function of magnetic field $B$ at $T = 0.7$ K (data from [97]). Such quantum oscillations were first observed in the resistance of YBCO at $p = 0.10$ [64] (see **Figure 10**). They come from electron orbits around a small closed pocket in the Fermi surface of this underdoped cuprate in its ground state, once superconductivity has been suppressed by the field.

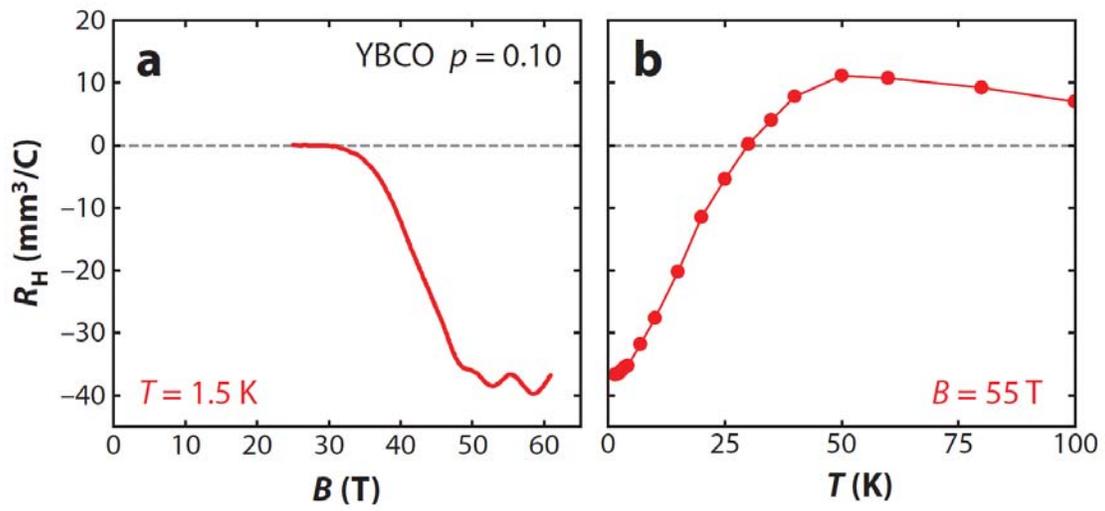

**Figure 10. Hall coefficient of YBCO.**

Hall coefficient of YBCO at $p = 0.1$, as a function of magnetic field $B$ at $T = 1.5$ K (**a**) and as a function of temperature at $B = 55$ T (**b**) (data from [68]). The fact that quantum oscillations are observed on a large negative background implies that they arise from orbits around a closed electron-like Fermi surface pocket, as confirmed by a negative Seebeck coefficient [41]. Adapted from [42].

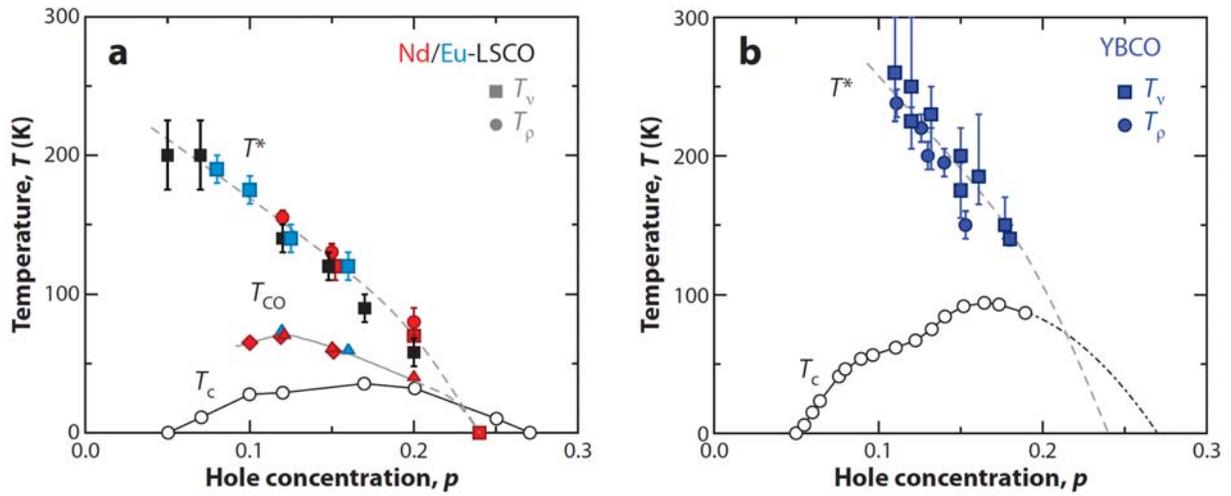

**Figure 11. Pseudogap temperature $T^*$ in LSCO and YBCO.**

Doping dependence of $T_\rho$ and $T_\nu$, the temperatures below which the resistivity $\rho$ and the Nernst coefficient $\nu$ respectively deviate from their linear behaviour at high temperature, two measures of the pseudogap crossover temperature $T^*$. **a)** (from [77]): $T_\rho$ for Nd-LSCO (red circles; see **Figure 8**); $T_\nu$ for Nd-LSCO (red squares; from [46]), Eu-LSCO (blue squares; from [46, 77]) and LSCO (black squares; obtained in [77] from data in [98]). Also shown are the superconducting temperature $T_c$ of LSCO (open black circles; from [98]) and the onset of stripe order in Nd/Eu-LSCO at $T_{CO}$ (triangles for nuclear quadrupole resonance, diamonds for X-ray diffraction; see **Figure 8**). **b)** (from [76]): equivalent data for YBCO, with $T_c$ data from [70]. The dashed line is a guide to the eye, and is the same dashed line as in panel **a**, multiplied by a factor 1.5.

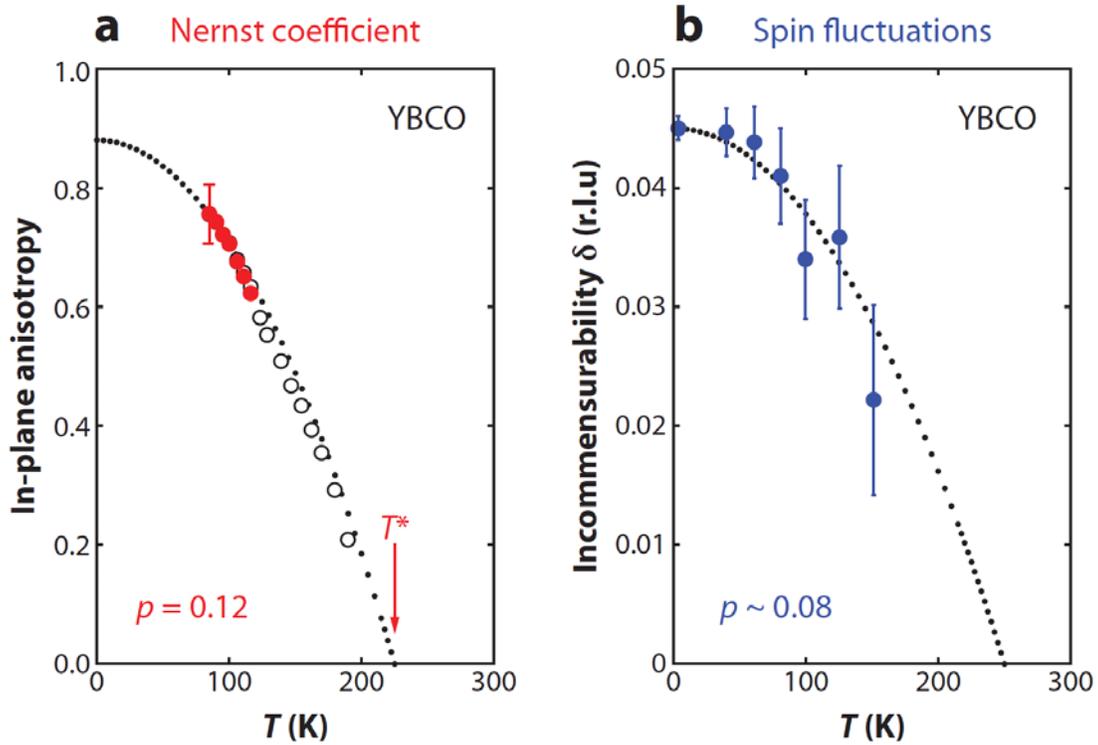

**Figure 12. Broken rotational symmetry in YBCO.**

**a)** *a-b* anisotropy ratio of the Nernst coefficient ν of YBCO at $p = 0.12$, plotted as $(\nu_b - \nu_a) / (\nu_b + \nu_a)$ (red dots) and $[D(T) - D(T_\nu)] / [S(T) - S(T_\nu)]$ (open circles) vs $T$, where $D(T) \equiv (\nu_a - \nu_b) / T$ and $S(T) \equiv - (\nu_a + \nu_b) / T$ (from [76]). The anisotropy grows with decreasing temperature, starting right at the pseudogap temperature $T^*$ (defined from $T_\nu$) for all dopings (see [76]). It shows that the pseudogap phase breaks the rotational symmetry of the $CuO_2$ planes [76]. **b)** Incommensurability (in reciprocal lattice units) of the spin fluctuation spectrum measured in YBCO at $p \approx 0.08$ by inelastic neutron scattering (data from [62]). This incommensurability is anisotropic, observed only along the $a^*$ axis. It reveals the appearance of unidirectional character in the SDW fluctuations below $T^*$. The dotted lines in both panels are a guide to the eye.